\documentclass[twocolumn,showpacs,preprintnumbers,amsmath,amssymb]{revtex4}

\usepackage{graphicx}
\usepackage{dcolumn}
\usepackage{bm}

\newcommand{\Eq}[1]{Eq.~({\protect\ref{#1}})}
\newcommand{\Fig}[1]{Fig.~\protect\ref{#1}}
\newcommand{\Table}[1]{Table {\protect\ref{#1}}}
\newcommand{\Ndf}{N_{\rm{DF}}}
\newcommand{\Nsmear}{N_{\rm{smr}}}
\newcommand{\Nconf}{N_{\rm{conf}}}

\newlength{\Tatescale}
\newcommand{\Hs}{\hspace*{1em}}

\newcommand{\BAR}[1]{\overline{#1}}

\newlength{\figwidth}
\begin{document}
\title{Scalar Glueball Mass Reduction at Finite Temperature \\
in SU(3) Anisotropic Lattice QCD}
\author{Noriyoshi ISHII}
\address{
The Institute of Physical and Chemical Research(RIKEN),\\
2-1 Hirosawa, Wako, Saitama 351-0198, JAPAN}

\author{Hideo SUGANUMA}
\address{Tokyo Institute of Technology,\\
2-12-1 Ohkayama, Meguro, Tokyo 152-8552, JAPAN}

\author{Hideo MATSUFURU}
\address{Yukawa Institute for Theoretical Physics, Kyoto University,\\
Kitashirakawa-Oiwake, Sakyo, Kyoto 606-8502, JAPAN}

\begin{abstract}
We  report  the  first  study  of  the  glueball  properties  at  finite
temperatures  below  $T_c$  using  SU(3) anisotropic  lattice  QCD  with
$\beta=6.25$, the  renormalized anisotropy $\xi \equiv a_s/a_t  = 4$ and
$20^3  \times N_t$  ($N_t$=35,36,37,38,40,43,45,50,72)  at the  quenched
level.  From the temporal correlation analysis with the smearing method,
about 20 \% mass reduction is observed for the lowest scalar glueball as
$m_G(T)=1250 \pm 50$MeV for $0.8 T_c  < T < T_c$ in comparison with $m_G
\simeq \mbox{1500$\sim$1700}$MeV at $T\simeq 0$.
\end{abstract}
\maketitle

Finite temperature QCD, including  the quark gluon plasma (QGP) physics,
is  one of the  most interesting  subjects in  the quark  hadron physics
\cite{qcd,miyamura,satz}.  At  high temperature, in  accordance with the
asymptotic  freedom of  QCD,  the strong  interaction  among quarks  and
gluons  is   expected  to  be   reduced,  and  there  would   occur  the
deconfinement and/or chiral  phase transition
\cite{qcd}.

For the  study of  finite temperature QCD,  the lattice QCD  Monte Carlo
simulation  provides  a reliable  method  directly  based  on QCD.   For
instance, SU(3)  lattice QCD  simulations at the  quenched level  show a
weak  first-order   deconfinement  phase  transition   at  the  critical
temperature  $T_c  \simeq 260$MeV  \cite{karsch1},  and  full SU(3)  QCD
simulations show a chiral phase transition at $T_c = 173(8)$MeV for $N_f
= 2$ and $154(8)$MeV for $N_f  = 3$ in the chiral limit \cite{full.qcd}.
Above  $T_c$,  most of  the  nonperturbative  properties  such as  color
confinement  and  spontaneous  chiral-symmetry breaking  disappear,  and
quarks and gluons are liberated.

Even below $T_c$, there are many  model predictions on the change of the
hadron  properties\cite{miyamura,hatsuda-kunihiro,ichie},  the mass  and
the   size,   due  to   the   change   in   the  inter-quark   potential
\cite{matsufuru,karsch2}  and  the  partial  chiral restoration.   As  a
pre-critical phenomenon of the QCD phase transition, the possible hadron
mass shift at the finite temperature  or in the finite density is now one
of  the  most interesting  subjects  in  hadron  and QGP  physics.   For
instance, the CERES data with the ultra-relativistic heavy-ion collision
experiment  may indicate  the $\rho$-meson  mass  shift\cite{ceres}, and
many theoretical  studies\cite{kapusta} have  been done to  explain this
experiment.

Nevertheless, lattice QCD studies  for thermal properties of hadrons are
still inadequate at  present because of the difficulty  in measuring the
hadronic  two-point correlators  on the  lattice at  finite temperature.
For instance, on the screening-mass measurement \cite{detar-kogut}, this
difficulty is due  to the mixture of the  large Matsubara frequencies in
addition  to the  absence  of technical  prescriptions  as the  smearing
method.  On  the other hand, on  the pole-mass measurement,  while it is
free from  the mixture of the Matsubara  frequencies, another difficulty
arises  from the  shrink of  the physical  temporal size  $1/T$  at high
temperature.  In  fact, the pole-mass measurements have  to be performed
within the limited distance shorter than $1/(2T)$, and such a limitation
corresponds to  $N_t = 4  \sim 8$ near  $T_c$ in the  ordinary isotropic
lattice QCD \cite{karsch1}.

To  avoid this  severe  limitation on  the  temporal size,  we adopt  an
anisotropic lattice where the  temporal lattice spacing $a_t$ is smaller
than the spatial  one $a_s$ \cite{matsufuru,klassen,taro,australia}.  We
can thus efficiently  use a large number of  the temporal lattice points
as $N_t  \sim 32$ even near  $T_c$, while the physical  temporal size is
kept fixed $1/T = N_t a_t$.
In this way, the number of available temporal data is largely increased,
and accurate pole-mass measurements from the temporal correlation become
possible \cite{taro,australia}.

In  this paper, we  study the  glueball at  finite temperature  from the
temporal correlation  analysis. We use SU(3) anisotropic  lattice QCD at
the  quenched level,  as a  necessary  first step  before attempting  to
include the  effects of  dynamical quarks in  the future.   Even without
dynamical  quarks, quenched  QCD can  reproduce well  various  masses of
hadrons,  mesons and baryons,  and important  nonperturbative quantities
such as the confining force and the chiral condensate.  In quenched QCD,
unlike full  QCD, the elementary  excitations are only glueballs  in the
confinement phase  below $T_c \simeq 260$MeV.  At  zero temperature, the
lightest physical  excitation is a scalar  glueball with $J^{PC}=0^{++}$
with  the mass  $m_G  \simeq \mbox{1500$\sim$1700}$MeV  \cite{australia,
morningstar,  weingarten,  teper}, which  is  expected  to dominate  the
thermodynamical properties below $T_c$.

We  consider  the   glueball  correlator  \cite{australia,  morningstar,
weingarten, teper, montvey, rothe} in  SU(3) lattice QCD as $G(t) \equiv
\langle \tilde  O(t) \tilde  O(0) \rangle$, $\tilde  O(t) \equiv  O(t) -
\langle O  \rangle$, $\displaystyle  O(t) \equiv \sum_{\vec  x} O(t,\vec
x)$.   The summation over  $\vec x$  physically means  the zero-momentum
projection.  The glueball operator $O(t,\vec x)$ is to be properly taken
so  as to reproduce  its quantum  number $J^{\rm  PC}$ in  the continuum
limit.  For  instance, the simplest composition for  the scalar glueball
is given as $O(t,\vec x)  \equiv \mbox{Re} \mbox{Tr} \{ P_{12}(t,\vec x)
+ P_{23}(t,\vec  x) + P_{31}(t,\vec x) \}$,  where $P_{\mu\nu}(t,\vec x)
\in$ SU(3) denotes the plaquette operator.
With the spectral representation, $G(t)$ is expressed as
$G(t)/G(0) = \sum c_n e^{-E_n t}$, $c_n  \equiv | \langle n | \tilde O |
0 \rangle |^2/  G(0)$, $G(0) = \sum |  \langle n | \tilde O  | 0 \rangle
|^2$,
where $E_n$ denotes the energy  of the $n$-th excited state $|n\rangle$.
Here,  $|0\rangle$  denotes  the  vacuum, and  $|1\rangle$  denotes  the
ground-state glueball.   Note that $c_n$  is a non-negative  number with
$\sum c_n  = 1$.   On a fine  lattice with  the spacing $a$,  the simple
plaquette  operator $P_{ij}(t,\vec  x)$  has a  small  overlap with  the
glueball ground  state $|G\rangle  \equiv |1\rangle$, and  the extracted
mass looks heavier owing to the excited-state contamination.  This small
overlap  problem originates  from  the  fact that  $O(t,\vec  x)$ has  a
smaller  ``size'' of  $O(a)$  than  the physical  peculiar  size of  the
glueball.  This problem  becomes severer as $a \to 0$.   We thus have to
improve $O(t,\vec x)$ so as to have almost the same size as the physical
size of the glueball.

One  of the  systematic  ways to  achieve  this is  the smearing  method
\cite{rothe,ape,takahashi}.   The smearing  method is  expressed  as the
iterative replacement of the original spatial link variables $U_i(s)$ by
the associated fat link  variables, $\BAR{U}_i(s) \in {\rm SU(3)}_c$, 
which is defined so as to maximize
\begin{equation}
	\mbox{Re}
	\mbox{Tr}
	\biggl[
		\BAR{U}_i^{\dagger}(s)
		\biggl(
			\alpha U_i(s)
		+	\sum_{j\neq i,\pm}
			U_{\pm j}(s)
			U_i(s\pm\hat j)
			U_{\pm j}^{\dagger}(s+\hat i)
		\biggr)
	\biggr],
\label{fat.link}
\end{equation}
where $U_{-\mu}(s) \equiv  U^{\dagger}_{\mu}(s - \hat\mu)$, and $\alpha$
is a real parameter. Here, the  summation is taken only over the spatial
direction  to   avoid  the  nonlocal  temporal   extension.   Note  that
$\BAR{U}_i(s)$  holds  the  same  gauge transformation  properties  with
$U_i(s)$.   We refer to  the fat  link defined  in \Eq{fat.link}  as the
first fat  link $U^{(1)}_i(s)$.  The  $n$-th fat link  $U^{(n)}_i(s)$ is
defined   iteratively  as  $U_i^{(n)}(s)   \equiv  \BAR{U}_i^{(n-1)}(s)$
staring from $U_i^{(1)}  \equiv \BAR{U}_i(s)$ \cite{takahashi}.  For the
physically  extended  glueball  operator,  we  use  the  $n$-th  smeared
operator, the plaquette operator constructed with $\BAR{U}_i^{(n)}(s)$.

The  smeared operator  physically corresponds  to an  extended composite
operator with  the original field  variable as $U_\mu(s)$.   We consider
the size of  the $n$-th smeared operator in terms  of the original field
variable.  
Using  the linearization  on the  gluon field,  we obtain  the diffusion
equation as \cite{australia,preparation}
\begin{equation}
        {\partial\over \partial n}
        K(\vec x, n)
=
        D \triangle K(\vec x, n),
\Hs
        D \equiv {a_s^2\over \alpha  + 4} 
\end{equation}
for the  distribution $K(\vec x,  n)$ of the  gluon field in  the $n$-th
smeared  plaquette, in  the case  of the  small spatial  lattice spacing
$a_s$.  The  $n$-th smeared  plaquette located at  the origin $\vec  x =
\vec 0$ physically correspond to the Gaussian extended operator with the
distribution as \cite{australia,preparation}
\begin{equation}
        K(\vec x, n)
=
        {1\over (\pi \rho^2)^{3/2}}
        \exp\left[ - {\vec x^2 \over \rho^2} \right], 
\end{equation}
where  $\rho$  represents  the   characteristic  size  of  the  Gaussian
distribution, and is defined as
\begin{equation}
        \rho
\equiv
        2\sqrt{D n} = 2 a_s \sqrt{n \over \alpha + 4}.
\label{size}
\end{equation}
Thus, the smearing method, which is introduced to carry out the accurate
mass  measurement by maximizing  the ground-state  overlap, can  be also
used to give  a rough estimate of the physical  glueball size.  In fact,
once  we obtain  the maximum  overlap with  some $n$  and  $\alpha$, the
glueball size is roughly estimated with \Eq{size}.

\begin{table}
\caption{The lattice QCD  result for the lowest scalar  glueball mass at
finite temperature.  The temporal  lattice size $N_t$, the corresponding
temperature $T$, the lowest scalar  glueball mass $m_G (T)$, the maximal
ground-state overlap $C^{\rm  max}$, fully correlated $\chi^2/\Ndf$, the
smearing number  $\Nsmear$, the number of  gauge configurations $\Nconf$
and  the 
rough estimate of the glueball size $\rho$ are
listed.  The most suitable  smearing number $\Nsmear$ is determined with
the maximum ground-state overlap condition.}
\label{table}
\begin{ruledtabular}
\begin{tabular}{cccccccc}
$N_t$ &
$T$[MeV] &
$m_G$[MeV] &
$C^{\rm max}$ &
$\chi^2/\Ndf$ &
$\Nsmear$ &
$\Nconf$ &
$\rho$[fm]\\
\hline
72 & 130 & 1450(40)	& 0.93(2) & 1.43 & 39 & 5541 & 0.42 \\
50 & 187 & 1410(46)	& 0.92(3) & 0.34 & 41 & 5168 & 0.44 \\
45 & 208 & 1456(34)	& 0.96(1) & 0.72 & 40 & 5929 & 0.43 \\
43 & 218 & 1323(39)	& 0.89(2) & 0.90 & 43 & 8693 & 0.45 \\
40 & 234 & 1260(45)	& 0.84(3) & 0.75 & 42 & 7420 & 0.44 \\
38 & 246 & 1221(35)	& 0.85(2) & 0.12 & 40 & 8736 & 0.43 \\
37 & 253 & 1273(32)	& 0.88(2) & 1.61 & 38 & 8633 & 0.42 \\
36 & 260 & 1208(35)	& 0.84(2) & 1.34 & 39 & 8603 & 0.42 \\
35 & 268 & 1188(34)	& 0.84(2) & 1.80 & 40 & 8462 & 0.43 \\
\end{tabular}
\end{ruledtabular}
\end{table}
We use the SU(3) anisotropic lattice plaquette action
\begin{eqnarray}
	S_G
&=&
	{\beta \over N_c}
	{1 \over \gamma_{\rm G}}
	\sum_{s, i<j \le 3}
	\mbox{Re}\mbox{Tr}
	\left( 1 - P_{ij}(s) \right)
\\\nonumber
&&
	+
	{\beta \over N_c}
	\gamma_{\rm G}
	\sum_{s, i \le 3}
	\mbox{Re}\mbox{Tr}
	\left( 1 - P_{i4}(s) \right)
\end{eqnarray}
with  the  plaquette operator  $P_{\mu\nu}(s)\in$  SU(3)  in the  $(\mu,
\nu)$-plane.   The   lattice  parameter   is  fixed  as   $\beta  \equiv
{2N_c}/{g^2}=6.25$,  and  the  bare  anisotropy parameter  is  taken  as
$\gamma_{\rm G} = 3.2552$ so as to reproduce the renormalized anisotropy
$\xi  \equiv a_s/a_t=4$  \cite{klassen}.  These  parameters  produce the
spatial lattice spacing as  $a_s^{-1}=2.341(16)$ GeV ($a_s \simeq 0.084$
fm),  and the  temporal  one as  $a_t^{-1}=9.365(66)$  GeV ($a_t  \simeq
0.021$ fm).  Here, the scale  unit is determined by adjusting the string
tension as  $\sqrt{\sigma}=440$MeV from the  on-axis data of  the static
inter-quark potential.  The pseudo-heat-bath algorithm is used to update
the gauge field configurations on  the lattice of the sizes $20^3 \times
N_t$, with  $N_t =  35, 36, 37,  38, 40,  43, 45, 50,  72$ as  listed in
\Table{table}.    For  each   temperature,  we   pick  up   gauge  field
configurations every  100 sweeps  for measurements, after  skipping more
than  20,000  sweeps  of  the  thermalization.   The  numbers  of  gauge
configurations used in our calculations are summarized in \Table{table}.

For completeness, we give an estimate of the critical temperature $T_c$.
To this end, we analyze the scattering plot of the Polyakov loop $P(\vec
x) \equiv {\rm Tr} \{ U_4(\vec x,0) \cdots U_4(\vec x,N_t-1) \}$ at each
gauge  field  configuration.   From  this analysis,  the  $\mathbb{Z}_3$
symmetry  holds at  $N_t=35$,  and the  system  is found  to  be in  the
confinement  phase.  On the  other hand,  the $\mathbb{Z}_3$  symmetry is
broken at $N_t=34$, which  indicates the deconfinement phase.  Hence, we
estimate  $T_c \simeq  270$MeV, which  is consistent  with  the previous
studies \cite{karsch1,matsufuru}.

We present the numerical results in SU(3) anisotropic lattice QCD at the
quenched level.  To enhance  the ground-state contribution, we adopt the
smearing method with the  smearing parameter $\alpha=2.1$, which we find
one of  the most suitable values  from the numerical  tests with various
$\alpha$.  The  statistical  errors  are estimated  with  the  jackknife
analysis \cite{montvey}.

\begin{figure}
\leftline{(a)}
\includegraphics[width=\figwidth]{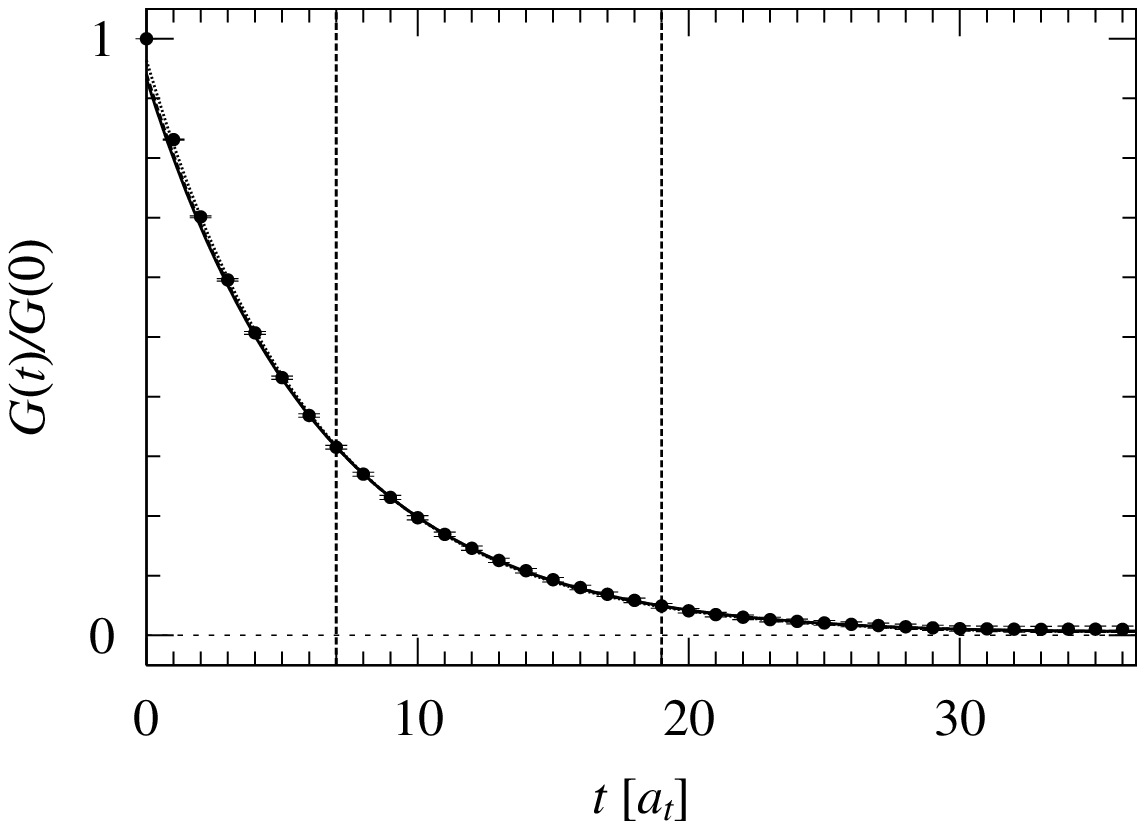}
\leftline{(b)}
\includegraphics[width=\figwidth]{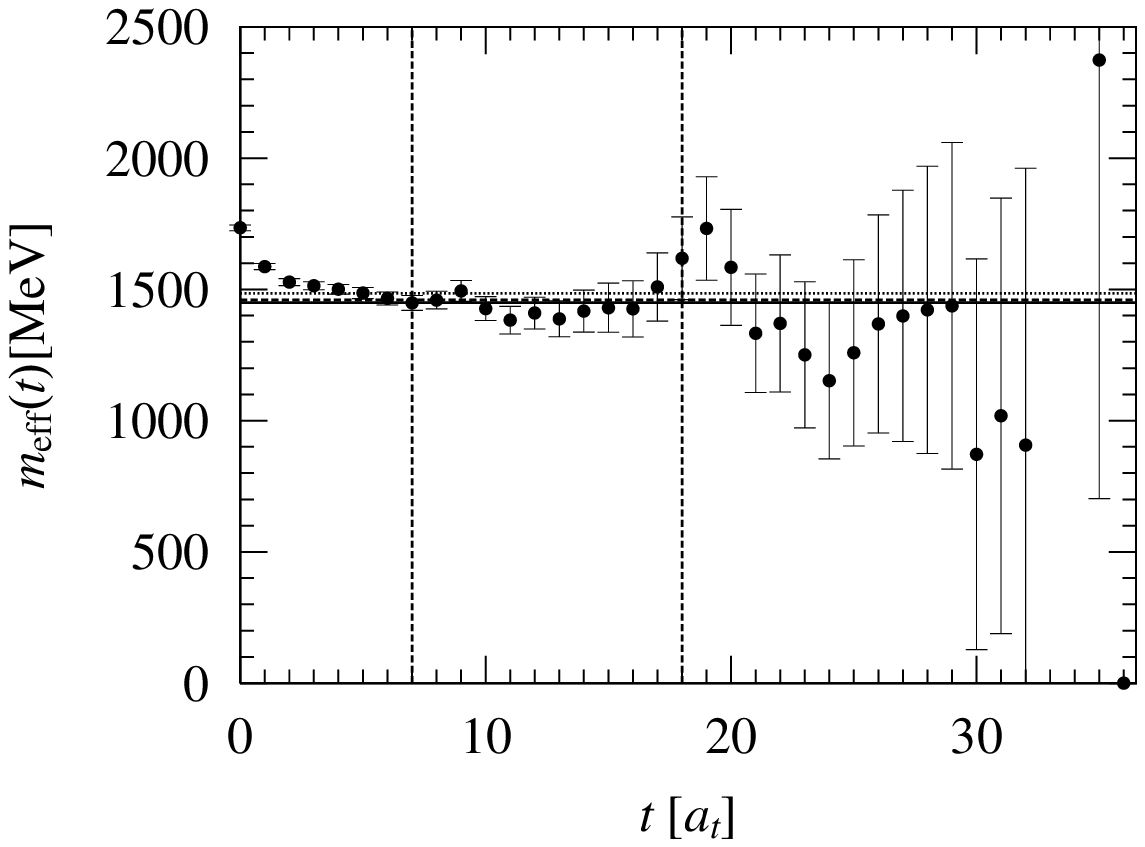}
\caption{(a) The  scalar glueball correlator $G(t)/G(0)$  for $\Nsmear =
40$ at  a low temperature  $T=130$MeV.  (b) The  corresponding effective
mass  plot. The  statistical  errors are  estimated  with the  jackknife
analysis.  
The  solid line denotes  the best  single hyperbolic  cosine fit  to the
lattice data in  the interval $[t_{\rm min}, t_{\rm  max}]$ indicated by
the two  vertical dashed  lines.  The dashed  and dotted curves  are the
best hyperbolic  cosine curves for  the modified fit range  with $t_{\rm
min}+1$ and  $t_{\rm min}+2$, respectively.  The closeness  of the three
curves means small fit-range dependence. 
}
\label{correlator.72}
\end{figure}
In  \Fig{correlator.72}(a),   we  show  a   scalar  glueball  correlator
$G(t)/G(0)$  at a  low temperature  $T=130$MeV for  the  smearing number
$\Nsmear=40$, where  most of the lattice  QCD data are well  fitted by a
single hyperbolic cosine, denoted by the solid curve, as
\begin{equation}
	G(t)/G(0)
=
	C(e^{-m_G t a_t} +  e^{-m_G (N_t -  t)a_t}).
\label{single.cosh}
\end{equation}
This indicates the achievement  of the ground-state enhancement owing to
the smearing method, and  then the excited-state contamination is almost
removed.

In general,  $G(t)/G(0)$ is  expressed as a  weighted sum  of hyperbolic
cosines  with  non-negative  weights,  and  $G(t)/G(0)$  decreases  more
rapidly   than  \Eq{single.cosh}   near  $t=0$   due   to  excited-state
contributions.  Hence,  $C$ should  satisfy  the  inequality  $C \le  (1
+e^{-m_G a_t  N_t})^{-1} \simeq 1$.  In the  ground-state dominant case,
$G(t)/G(0)$ can be well approximated  by a single hyperbolic cosine, and
$C \simeq 1$ is realized.  We refer to $C$ as the ground-state overlap.

From  \Fig{correlator.72}(a),  we find  $C  \simeq  1$  and $m_G  \simeq
1450$MeV for the lowest scalar glueball mass at a low temperature.  This
seems consistent with $m_G \simeq \mbox{1500$\sim$1700}$MeV at $T \simeq
0$ \cite{morningstar,weingarten,teper}.

\begin{figure}
\leftline{(a)}
\includegraphics[width=\figwidth]{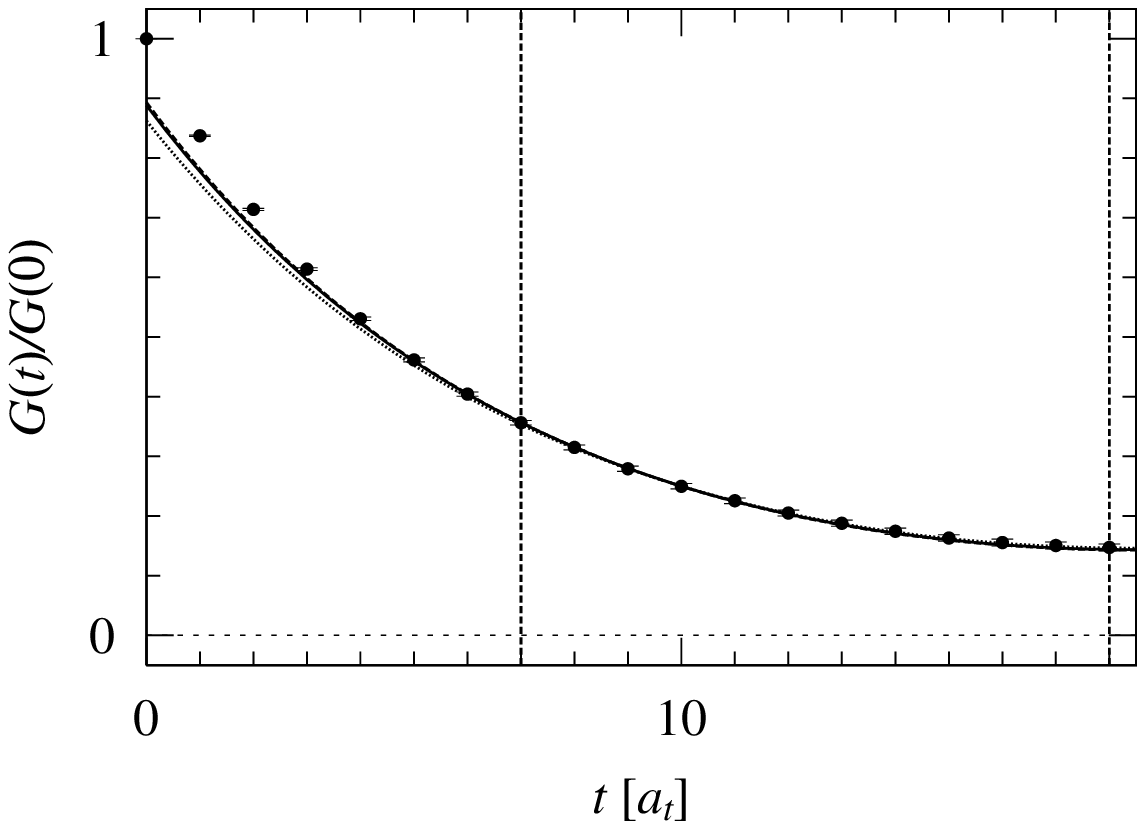}
\leftline{(b)}
\includegraphics[width=\figwidth]{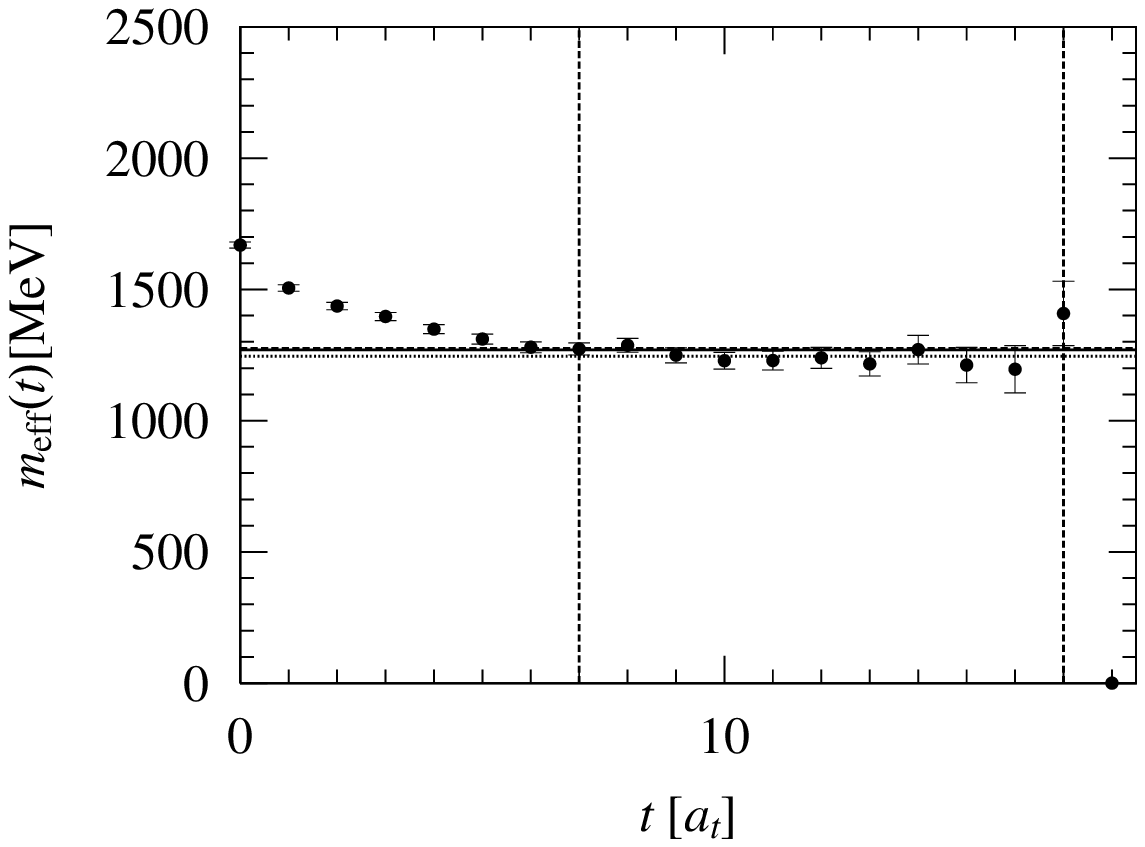}
\caption{(a) The  scalar glueball correlator $G(t)/G(0)$  for $\Nsmear =
40$ at  a high temperature  $T=253$MeV. (b) The  corresponding effective
mass  plot.  The  statistical errors  are estimated  with  the jackknife
analysis.  
The  solid line denotes  the best  single hyperbolic  cosine fit  to the
lattice data in  the interval $[t_{\rm min}, t_{\rm  max}]$ indicated by
the two  vertical dashed  lines.  The dashed  and dotted curves  are the
best hyperbolic  cosine curves for  the modified fit range  with $t_{\rm
min}+1$ and  $t_{\rm min}+2$, respectively.  The closeness  of the three
curves means small fit-range dependence.  }
\label{correlator.37}
\end{figure}
In  \Fig{correlator.37}(a),   we  show  a   scalar  glueball  correlator
$G(t)/G(0)$  at a high  temperature $T=253$MeV  for the  smearing number
$\Nsmear=40$. Owing to a suitable smearing, most of the lattice QCD data
are  well fitted  by a  single hyperbolic  cosine denoted  by  the solid
curve.

Each  best fit  analysis is  performed  in the  interval $[t_{\rm  min},
t_{\rm  max}]$,  which  is  determined  from the  flat  region  $[t_{\rm
min},t_{\rm max} - 1]$  appeared in the corresponding ``effective mass''
plot  shown in Figs.~\ref{correlator.72}(b)  and \ref{correlator.37}(b).
The effective mass $m_{\rm eff}(t)$ is a solution of
\begin{equation}
{G(t+1)\over G(t)}
=
{
	\cosh\left( m_{\rm eff}(t)a_t(t + 1 - N_t/2) \right)
\over
	\cosh\left( m_{\rm eff}(t)a_t(t - N_t/2) \right),
}
\end{equation}
for a given $G(t+1)/G(t)$ at  each fixed $t$ \cite{montvey}.  
In Figs.~\ref{correlator.72}  and \ref{correlator.37}, we  show also the
results  of  further two  fits  in  the  modified interval  as  $[t_{\rm
min}+1,t_{\rm max}]$  and $[t_{\rm  min}+2,t_{\rm max}]$ by  dashed line
and  dotted  line, respectively.   The  closeness  of  the three  curves
suggests small fit-range dependence.

In the most suitable smearing $\Nsmear$, the ground-state overlap $C$ is
maximized  and  the  mass   $m_G$  is  minimized,  which  indicates  the
achievement  of  the  ground-state  enhancement.  (For  extremely  large
$\Nsmear$,  the  operator  size  exceeds  the  physical  glueball  size,
resulting  in   the  decrease  of   the  overlap  $C$.)    In  practical
calculations, the maximum overlap and the mass minimization are achieved
at almost  the same  $\Nsmear$, and both  of these two  conditions would
work as an indication of the maximal ground-state enhancement.  Here, we
take the maximum  ground-state overlap condition as $C  \simeq 1$.  (The
mass  minimization condition  leads  to almost  the  same glueball  mass
\cite{preparation}.)

\begin{figure}
\includegraphics[width=\figwidth]{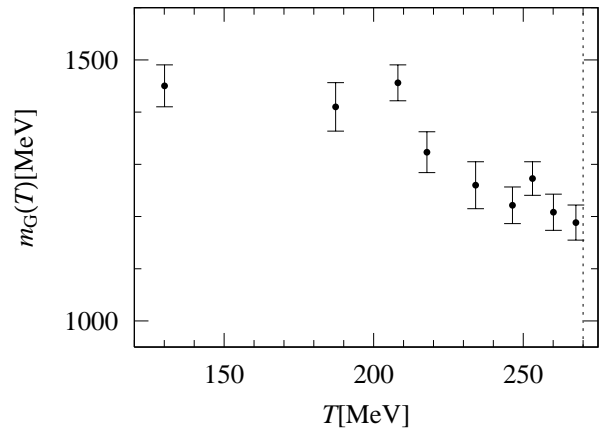}
\caption{The lowest scalar glueball mass plotted against the temperature
$T$.  It is obtained with the best hyperbolic cosine fit in the interval
$[t_{\rm  min}, t_{\rm  max}]$ determined  from the  flat region  in the
effective  mass plot. The  vertical dotted  line indicates  $T_c\simeq $
260MeV.  }
\label{mass.reduction}
\end{figure}
From  the analysis  at various  temperatures,  we plot  the lattice  QCD
result for the lowest scalar glueball mass $m_G (T)$ against temperature
$T$ in \Fig{mass.reduction}.  We observe, in \Fig{mass.reduction}, about
20 \% mass  reduction or a few hundred MeV mass  reduction of the lowest
scalar glueball near $T_c$ as $m_G(T)=1250  \pm 50$MeV for $0.8 T_c <T <
T_c$ in comparison with  $m_G(T \sim 0) \simeq \mbox{1500$\sim$1700}$MeV
\cite{morningstar,weingarten,teper}.

We also  give a  rough estimate  of the glueball  size. To  estimate the
glueball  size,   we  search   $\Nsmear$  which  realizes   the  maximum
ground-state overlap $C^{\rm max}$.  From \Eq{size} with this $\Nsmear$,
we roughly estimate the glueball size  as $\rho \simeq 0.4 \sim 0.45$ fm
both at low temperature and at high temperature near $T_c$. Thus, we see
that the thermal effect on the  glueball size is rather small, which may
provide an  important information in the  bag model argument  of the QCD
phase transition \cite{australia,preparation}.


In \Table{table}, we summarize the lowest scalar glueball mass $m_G(T)$,
the ground-state overlap  $C^{\rm max}$, fully correlated $\chi^2/\Ndf$,
the  corresponding  smearing  number  $\Nsmear$,  the  number  of  gauge
configurations $\Nconf$ and the estimated glueball size $\rho$.

Thus,  the present  lattice QCD  calculation indicates  that  the lowest
scalar glueball exhibits about  250MeV mass reduction near $T_c$ keeping
its size.   Here, we  briefly discuss the  physical consequence  of this
result,  considering  the  trigger  of  the QCD  phase  transition.   In
quenched QCD below $T_c$, the  lowest glueball is the lightest particle,
and  its thermal  excitation is  expected to  have primary  relevance at
finite temperature.   However, lattice QCD indicates $m_{\rm  G} > 1$GeV
even near  $T_c$, and therefore the thermodynamical  contribution of the
glueball seems  strongly suppressed by  the small statistical  factor as
$e^{-m_{\rm   G}/T}$   near   $T_c   \simeq   260$MeV   \cite{australia,
preparation}.  This  may indicate  that the thermal  glueball excitation
does not play  the relevant role in the  deconfinement phase transition,
at least in quenched QCD.  Then,  what is the driving force to bring the
phase  transition  ?   In  this  way,  our  result  brings  up  such  an
interesting new puzzle on the QCD phase transition.
%

Several comments are in order.  The first comment is on the closeness of
our simulations  to the  continuum limit.  
%
%
In   Ref.\cite{continuum.limit},   the   authors  investigated   $\beta$
dependence  of glueball masses  at zero  temperature, and  estimated the
discretization error on the scalar glueball mass to be less than 5 \% at
$\beta=6.4$.  According  to them, the discretization  error is estimated
about 6 \% at $\beta = 6.25$ in the present calculation.
The  second comment  is  on the  finite  volume artifact  on the  scalar
glueball mass.   In Ref.\cite{morningstar}, Monte  Carlo simulations 
were performed  on the lattice of the  physical size $(1.76{\rm{fm}})^3$
and $(1.32{\rm{fm}})^3$  at zero  temperature to investigate  the finite
volume  errors in  the  various  glueball masses  by  using an  improved
action.   The authors  concluded  that the  systematic error  in the
lowest scalar glueball mass from the finite volume is negligible at zero
temperature.   Note  that  the  finite  volume artifact  on  the  scalar
glueball mass  is essentially independent of  the regularization method,
i.e.,  a  specific  choice  of   the  lattice  action,  as  far  as  the
discretization  is enough  fine.   It follows  that,  the finite  volume
artifact of our  results are negligible, since the  physical size of our
lattice is $(1.68{\rm fm})^3$.

To  summarize,  we  have  studied  the  glueball  properties  at  finite
temperature   using  SU(3)   anisotropic  quenched   lattice   QCD  with
5,000--9,000  gauge  configurations   at  each  temperature.   From  the
temporal correlation analysis with the smearing method, we have observed
about 20\%  mass reduction  of the lowest  scalar glueball as  $m_G(T) =
1250 \pm 50$MeV  for $0.8 T_c < T <T_c$, while  no significant change is
seen for meson masses near $T_c$ in lattice QCD \cite{taro}.
%

Finally,  we  comment  the  brief  outlook.   It  seems  interesting  to
investigate  other glueballs  such as  the $2^{++}$  glueball  at finite
temperature to clarify whether the thermal mass reduction is peculiar to
the lowest scalar glueball or  universal feature in glueballs.  It would
be also interesting to analyze  the spectral function of the glueball at
finite temperature  from its temporal  correlation in terms of  the mass
and  the thermal  width, because  the width  broadening may  provide the
similar  effect to the  temporal correlator  \cite{kapusta} as  the mass
reduction.  Our result shows that  the scalar glueball mass reduction is
about  250MeV, which  is enough  large, and  therefore the  thermal mass
shift  of  the scalar  glueball  may  become  observable in  the  future
experiment in RHIC.

H.~S.  is supported by  Grant for Scientific Research (No.12640274) from
the  Ministry  of Education,  Culture,  Science  and Technology,  Japan.
H.~M. is  supported by  Japan Society for  the Promotion of  Science for
Young  Scientists.   The lattice  calculations  have  been performed  on
NEC-SX5 at Osaka University.

\end{document}